# Achieving Bright Organic Light Emitting Field Effect Transistors with Sustained Efficiency through Hybrid Contact Design

*Shih-Wei Chiu[1], An Hsu[1], Lei Ying[2], Yong-Kang Liaw[1], Kun-Ta Lin[1], Jrjeng Ruan[1], Ifor D. W. Samuel[3] and Ben B. Y. Hsu[1*]*

1 *Department of Materials Science and Engineering, National Cheng Kung University, Taiwan.*

2 *Institute of Polymer Optoelectronic Materials and Devices, State Key Laboratory of Luminescent Materials and Devices, South China University of Technology, Guangzhou 510640, P. R. China.*

3 *Organic Semiconductor Centre, SUPA, School of Physics and Astronomy, University of St Andrews, U.K.*

Keywords: light emitting transistors, low efficiency roll-off, high brightness, semiconducting conjugated polyelectrolytes, band bending

## Abstract

Organic light emitting field effect transistors (OLEFETs) with bilayer structures have been widely studied due to their potential to integrate high-mobility organic transistors and efficient organic light emitting diodes. However, these devices face a major challenge of imbalance charge transport leading to severe efficiency roll-off at high brightness. Here, we propose a solution to this challenge by introducing a transparent organic/inorganic hybrid contact with specially designed electronic structures. Our design aims to steadily accumulate the electrons injected to the emissive polymer, allowing the light emitting interface to effectively capture more holes even when hole current increases. Our numerical simulations show that the capture efficiency of these steady electrons will dominate charge recombination and lead to a sustained external quantum efficiency of 0.23% over 3 orders of magnitude of brightness (4 to 7700 cd/m$^2$) and current density (1.2 to 2700 mA/cm$^2$) from -4 to -100 V. The same enhancement is retained even after increasing EQE to 0.51%. The high and tunable brightness with stable efficiency offered by hybrid-contact OLEFETs make them ideal light emitting devices for various applications. These devices have the potential to revolutionize the field of organic electronics by overcoming the fundamental challenge of imbalance charge transport.

# Introduction

Organic light emitting field effect transistors (OLEFETs) are a promising new class of light emitting devices that combine the functionality of a switch for both electrical current and optical emission.[1–5] Their integration of an organic light emitting diode (OLED) and its driving circuit into a single device makes OLEFETs a convenient and versatile option for low-cost display and lighting technologies.[6–12] However, to meet the diverse demands of lighting and display applications, it is essential to have strong and controllable light intensity with stable external quantum efficiency (EQE).[13–16] Despite significant efforts to improve both emission intensity[10,17–21] and EQE[22–25], it remains a challenge to achieve high brightness with sustained EQE at large current density.

Sustaining high EQE in ambipolar light emitting materials requires a balanced electron and hole transport. To achieve the highest EQE in such OLEFETs, the charge transport must be delicately balanced at $|V_{GD}| = |V_{GS}|$ with equalized electron and hole currents in their minima. Any voltage deviation from $|V_{GD}| = |V_{GS}|$ will amplify either hole or electron current and decrease EQE. Hence the highest EQE only appears at a narrow current window around its minimum and limit the achievable optical intensity.[26–30] High-mobility OLEFETs, on the other hand, can provide high emission intensity but operating these hole-dominated transistors at elevated current results in imbalanced transport and degraded EQE. A recent study showed that high-mobility OLEFETs with enhanced electron injection can achieve an EQE of 0.8% at a brightness about 10,000 cd/m$^2$, but fell to 0.4% at the highest brightness 29,000 cd/m$^2$.[20] Another work demonstrated 0.1% EQE at a brightness around 10,000 cd/m$^2$ but dropped to 0.02% at the highest brightness around 20,000 cd/m$^2$.[10] For years, matching charge transports in OLEFETs has been the conventional method to enhance performance but that only achieves either high EQE or high intensity in narrow current ranges. A major challenge remains in sustaining high EQE over a wide range of current.

In this report, we focused on Coulombic capture and developed a novel method to enhance electron injection, charge capture, and optical emission simultaneously. This was accomplished by incorporating a transparent organic/inorganic hybrid contact that consists of a carefully designed electron-accumulating interface and a n-type conjugated polyelectrolyte (CPE). The use of n-type CPE facilitated electron conductivity and electronic alignment with the p-type light-emitting polymer, causing a band-bending effect. This leads to electrons accumulating in the potential well on the p-type polymer side, which is not achievable with conventional neutral or ionic CPEs.[31–33] The accumulated high-density electrons efficiently capture more holes and dominate charge recombination, resulting in very low efficiency roll-off. Our findings show that the maximum EQE degradation at high currents was 14−22%, compared to 2- and 5-fold degradation in previous high-intensity OLEFETs with the same emission polymer.[10,20] We have successfully developed a bi-functional OLEFET that integrate electric switch and light emission. Our device exhibits adjustable intensity, sustained efficiency, and a logic-circuit function.

## Results and Discussion

Figure 1a shows the complete device architecture and its actual picture. A pair of the n-type contact (drain) and its counter p-type contact (source), which forms a hybrid contact-OLEFET (HC-OLEFET), is introduced for more balanced charge injections. Poly(3,6-dialkylthieno[3,2-b]thiophene-co-bithiophene) (PATBT) is the p-type high-mobility semiconducting polymer. Poly[9,9-bis(6"-(*N*,*N*-dimethyl)-hexyl)-2,7-fluorene]-*alt*-[3,7-dibenzothiophene-*S*,*S*-dioxide] (SPFN) is the n-type CPE.[34,35] The source is composed in order of PATBT (30 nm)/Au (90 nm) and the drain is composed of SPFN (2 nm)/Au:Ag (1:7 nm)/MoO$_x$ (30 nm). The hybrid drain of this work consists of a bimetallic contact (Au and Ag) and SPFN which serve as the inorganic and organic parts. Super yellow (SY) is the p-type light emitting polymer. Their chemical structures are shown in Figure 1b. To prevent oxidation, a protection layer of MoO$_x$ was deposited. The polar amino side group of SPFN and its lowest unoccupied molecular orbitals (LUMO) within the large energy mismatch between SY and Ag can improve electron injection[34,35]. Figure 1c shows the schematic energy diagrams and the corresponding absorption spectra before (top) and after (bottom) SPFN contacts SY. By aligning the Fermi levels of the heterojunction through charge transfer between p-type SY and n-type SPFN,[36,37] $E_F^{SY}$ and $E_F^{SPFN}$, they will reach a new thermal equilibrium that causes a depletion of holes in SY HOMO. As the result, SY LUMO is bent downward, as shown in Figure 1c inset.[38,39] These changes enable a series of electronic transitions in SY, giving rise to a new spectral feature at 545 nm.[40,41] Upon biasing the HC-OLEFETs, electrons and holes will transport and accumulate at the SY interface (Figure 1d). These accumulated electrons have low mobility and long transit time, leading to enhanced capture efficiency in SY.

Understanding the origin of the band bending effect on the light emitting interface is critical in selecting appropriate electron injection materials. In the realm of n-type CPE, prioritizing electron conductivity as opposed to molecular dipoles of conventional ionic or neutral CPEs[32,52] is crucial to facilitate efficient charge transfer between p-type and n-type semiconductors. Neutral CPEs and small molecules with n-type moieties, such as the highly-conjugated SPFN,[42] Bphen,[13,14] TPBI,[15] and TmPyPB[16] can enhance electron conduction, leading to the band alignment as illustrated in Figure 1c. These neutral n-type organic semiconductors with effective electron conduction are preferred to avoid phenomena such as delayed transport, current hysteresis, and luminescence quenching reported in ionic CPEs.[43–45]

To maximize both optical emission and electrical conduction, it is essential to have an ultrathin hybrid drain (<< optical wavelength) with a smooth surface. As seen in Figure 2a, atomic force microscope images of 7 nm Ag without (left) and with (right) 1 nm Au demonstrate the impact of Au on surface smoothness. The Ag without Au forms larger particles with a rougher surface, causing significant light scattering. Image analysis (using ImageJ and NanoScope) shows that the average particle size is 10.0 nm with a surface roughness $R_q$= 1.3 nm for the drain without Au, while with Au, the average size is 4.0 nm with $R_q$= 0.7 nm (Figure S1 in the supporting information, SI).

The drain without Au appears bluish and opaque due to strong scattering, whereas with Au acting as a wetting layer, the drain looks transparent (inset of Figure S2a). The transmittance of the Au-added hybrid drain on glass substrates was measured to be about 80%, compatible to commercial ITO glasses in the wavelength range 550 to 630 nm. This wavelength range coincides with the maximum electroluminescence of SY (refer to Figure S2ab). Note that the transparency is only optimized for SY so the hybrid drain is semi-transparent in other wavelengths. There are more details about transmittance of the hybrid drains affected by SPFN molecular weight and $MoO_x$ thickness, provided in Experimental Methods.

The electronic performance of the hybrid drain strongly depends on its morphology. To reduce the energy mismatch for electrons between Ag and SY, a smooth SPFN layer was required. Otherwise, thick and rough SPFN will result in an injection barrier and light scattering (Please see the degraded transmittance and current in Figure S3 and S4 for the hybrid drain on a rough SPFN). The reduction in mismatch was measured using a Kelvin probe force microscope (KPFM). Prior to the measurement, the SPFN was smoothed by 3 rounds of methanol flushing to prevent jittering potential readings due to the collision between KPFM tip and aggregates (see Figure S5). The surface potential mapping images of SY and SY/SPFN on indium tin oxide (ITO) substrates are shown in Figure 2b, with uniform surface potentials read as $\Phi_{SY}$= -296 and $\Phi_{SY/SPFN}$= -85 mV (with $\Phi_{ITO}$ as reference). This resulted in a 211 mV reduction in mismatch, leading to a smaller barrier to improve electron conduction. Conductive AFM (cAFM) scans on the SY/SPFN sample after KPFM measurements showed an inverse dependence of topographic heights on methanol flushing times; both smoothness and currents from ITO to the Au-coated cAFM tip increased with flushing as shown by the red-circled areas Figure 2c. The I-V curves obtained by cAFM showed a 10-fold increase in conduction from 10 to 101 pA (refer to Figure S6). Because the deeper highest occupied molecular orbitals (HOMO) of SPFN compared to that of SY (Figure 1c) are particularly designed to block the hole current in SY, the improved conduction is attributed to electrons and is expected to enhance recombination current for light emission in HC-OLEFETs.

We studied the transfer and output I-V characteristics as well as the optical responses of two HC-OLEFETs. Figure 3a displays the transfer I-V curves obtained at a drain-to-source voltage $V_{DS}$= -100 V, scanned from 12 to -100 V for both devices. The transport mobilities, 0.23 cm$^2$/Vs and 0.52 cm$^2$/Vs, were extracted from the saturation regime of the transfer curves using the transconductance equation: $I_{DS} = \frac{WC}{2L}\mu(V_{GS} - V_{th})^2$; where W, L, and C are the channel width and length, and capacitance, respectively. $I_{DS}$, $V_{GS}$, $V_{th}$ are source-drain current, gate-to-source voltage, and threshold voltage. The extracted hole mobilities, on/off ratios, current densities on emission area, brightness, and EQE are collected in Table 1. The results indicate that both devices have higher brightness with increasing $I_{DS}$, and the device with higher mobility ($\mu$=0.52 cm$^2$/Vs) reaches the highest brightness. However, the output I-V curves (Figure 3b), obtained by scanning $V_{DS}$ from 0 to -100 V with varied $V_{GS}$, show that the brightness curves tend to roll-off at large current for the device with higher mobility device while the device with lower mobility ($\mu$=0.23 cm$^2$/Vs) approached saturation with limited roll-off. The more obvious roll-off in the high-mobility device can be explained by the larger EQE degradation 22% compared to that of 14% in the low mobility device, shown in Figure 4d. Thus, we conclude that the lower mobility device has higher efficiency.

Figure 3c illustrates the reliability of the device through an electroluminescence video filmed during an output I-V scan. The bright and uniform emission of light in the entire pixel area (even at large current $V_{GS}$= -20 V) demonstrates the absence of voltage loss along the electrodes. Unlike typical OLEFETs often part of pixels emits light. High-EQE ambipolar OLEFETs show wiggling[27,28] or scattering[29] pixels and high-brightness unipolar ones typically emit from a reduced area on opaque metal edge[19–21,33]. HC-OLEFETs show that high electron conductivity produces uniform light emission across all pixels. This is attributed to the ultrathin, smooth, and transparent hybrid contacts that provide well-defined and intact emission areas.

The hole-dominated HC-OLEFETs demonstrate a strong relationship between optical and electrical performance, as shown by taking the square root of $I_{DS}$ and brightness (denoted $SRI_{DS}$ and SRB) in Figure 3a. Figure 4ab show that the $SRI_{DS}$ and SRB vs. $V_{GS}$ curves for both devices are nearly identical, implying that the optical output obeys the transistor-based transconductance equation. The brightness-to-$I_{DS}$ curves in Figure 4c are linear, and the sustained EQE curves in Figure 4d show two stable values of about 0.51% for 0.23 cm$^2$/Vs and about 0.23% for 0.52 cm$^2$/Vs. These devices retain nearly constant EQE across a wide range of operating currents, with a drop of <22% at the largest current regimes, compared to a 2- to 5-fold degradation observed in similar works using SY.[10,20] Conversely, a HC-OLEFET composed of an ionic CPE fails to sustain EQE with increasing hole current and rapidly loses EQE at large current regimes, as shown in Figure S10.

We investigated the roll-off observed at the high $I_{DS}$ regime in the 0.52 cm$^2$/Vs device, which suggests a substantial efficiency degradation. The EQE equation (the equation S1 in the SI) reveals that the degradation at high current density can be caused by a decrease in photoluminescence quantum yield $\eta_{PL}$ (singlet quenching) and capture efficiency $\eta_c$ (imbalanced charges). Previous studies have shown that a current density of at least 6 A/cm$^2$ is needed to induce substantial singlet quenching by triplets and polarons.[46–48] As our current density is <2.7 A/cm$^2$, the efficiency roll-off is mainly due to imbalanced charge transport, dominated by holes. Conventional solutions for two-terminal OLEDs, such as slowing hole transport and accelerating electrons, have limited success in three-terminal OLEFETs with more complex electrical and optical responses. We will present a solution to maintain high electrical performance without degrading optical efficiency through electron accumulation.

To account for the observed stable EQE, we developed a model of capture dynamics based on the Langevin recombination to simulate capture efficiency. Since the efficiency roll-off is dominated by imbalanced charge transports, the interplay between electron/hole/capture currents in SY will determine the capture efficiency $\eta_c$ as the equation (1) shows.

$$\eta_c = \frac{J_c}{J_p + J_n - J_c} \tag{1}$$

where the electron and hole currents are $J_n = ne\mu_n E$ and $J_p = pe\mu_p E$. We adopted the Langevin theory for the capture current $J_c = eRl$ with a Langevin recombination rate $R = \gamma np = e(\mu_p + \mu_n)np/\epsilon_{SY}$ per second per volume and the mean capture distance $l$. Though the Langevin model is commonly used in disordered organic semiconductors like SY, it is only valid when charge mean free path $\lambda$ is much smaller than charge thermal capture radius $r_c$ of Coulomb potential. For SY at room temperature (25 °C) with dielectric constant $\epsilon_{SY} = 3.2\epsilon_o$, $r_c = e^2/(4\pi\epsilon_{SY}k_B T)$ is found to be 18 nm and the typical scale of hopping distance $\lambda$ around 1 nm; 1 nm $\ll$ 18 nm which justifies the model. In a poor conductor such as SY, slow charge transport is characterized by charge displacement and restoration, known as dielectric relaxation. The interactive distances of electrons and holes in motion can be determined through the dielectric response, which are defined as $l_n$ and $l_p$. Hence the equation (1) can be rewritten as

$$\eta_c = \frac{J_c/J_p}{1 + J_n/J_p - J_c/J_p} = \frac{(1 + \frac{\mu_n}{\mu_p}) \cdot \frac{l}{l_n}}{1 + \frac{n\mu_n}{p\mu_p} - (1 + \frac{\mu_n}{\mu_p}) \cdot \frac{l}{l_n}} \tag{2}$$

where the capture distance $l$ is approximated by the geometric mean of $l_n$ and $l_p$, $l = \beta l_n l_p / (l_n + l_p)$; $\beta$ is the effective interaction strength through dielectric response between charges, treated as a fitting output. Accumulated electron density n is another fitting output; the maximum electron density $n_{max}$, $10^{18}$ 1/cm$^3$, is estimated by assuming 100 % recombination, treated as the initial value in the simulation. Meanwhile, the hole density p is the fitting variable (a function of I$_{DS}$). The calculations for $l_n$, $l_p$, p, $p_{max}$, $n_{max}$, and $\eta_c$(n, p) are provided in the Experimental Methods.

In order to study capture dynamics, a series of mobility ratios $\mu_n/\mu_p$ were incorporated into the equation (2). The simulated $\eta_c$ curves were optimized using the least-squares method, and the fitting results (n and $\beta$) are summarized in Table 2. For device 2 with larger hole mobility, the obtained $\beta$ values are consistently lower than those extracted from device 1. This outcome suggests a relatively weaker interactive strength with electrons and a more pronounced roll-off in Figure 3b due to faster hole movement. The resulting $\eta_c$ curves are plotted in Figure 4e and 4f and show two distinct capture dynamics, represented by open and solid triangles for small and large $\mu_n/\mu_p$ regimes respectively. When $\mu_n/\mu_p$ is small, electrons are confined in the potential well, resulting in localized dielectric interactions and smaller $\beta$ values. The simulated $\eta_c$ curves were found to hold over a wide range of I$_{DS}$, indicating a well-controlled kinetic process. On the other hand, when $\mu_n/\mu_p$ is large, more electrons escape the potential well and into SY where abundant holes capture electrons quickly, resulting in diffusion-like transport. With increasing $\mu_n/\mu_p$, electrons can further diffuse into SY and leads to larger and more homogenous dielectric interactions with holes (all $\beta$ are shown around 4×10$^{-5}$), causing $\eta_c$ to gradually increase with rising I$_{DS}$. The two distinct regimes with varying interaction strengths $\beta$ indicate distinguishable dynamic behaviors based on the mobility ratios, revealing the existence of two capture mechanisms.

As $\mu_n/\mu_p$ increases, the fitted electron densities decrease and capture dynamics transition from one mechanism to another. At n≤1.4×10$^{16}$ 1/cm$^3$, the transitioning $\eta_c$ curves collapse into one, indicating identical capture dynamics. The conversion of n=1.4 × 10$^{16}$ 1/cm$^3$ into the diffusion length $\lambda_n$ =18 nm (as described in the Experimental Methods) reveals that the transport of escaped electrons is limited by the thermal capture radius of Coulomb forces (18 nm in our cases). Therefore, well-controlled and Coulomb force-controlled capture mechanisms are suggested as shown in Figure 5. The overall deviation of all $\eta_c$ curves in Figure 4e and 4f is only 14% and 22%, showing a remarkable improvement in capture efficiency despite imbalanced charge transport.

The modeling result reveals that high brightness with low efficiency roll-off is achievable by incorporating a high-density electron accumulation layer on the light emitting interface. The achieved high and tunable brightness with stable EQE and logic functions on a well-defined emission area in a single HC-OLEFET can meet the widest applications which cannot be satisfied by OLEDs.

## Conclusions

In conclusion, our study presents the development of a HC-OLEFET with transparent organic/inorganic hybrid contact on the emissive SY. The novel design composed of bimetallic contact and n-type SPFN enhances electron injection and induces band bending to accumulate high-density electrons at the SY/SPFN interface, leading to sustained EQE across a wide range of current levels. Our modeling of capture dynamics shows that the efficiency is dominated by Coulombic capture when electron mobility is high and by bending potential well when it is low. This breakthrough, achieved by carefully engineering the interfacial electronic structures, addresses the long-standing challenge of efficiency roll-off in OLEFETs. It can be applied to organic semiconductors with imbalanced charge transport and provide practical advancements toward high-efficiency display and lighting technologies.

## Experimental Methods

**Materials information.** Poly(3,6-dialkylthieno[3,2-b]thiophene-co-bithiophene) (PATBT, lisicon SP220) and poly(1,4-phenylenevinylene) PPV copolymer Super Yellow (SY, Livilux PDY-132) were both purchased from Merck KGaA. PATBT has number-average molecular weight ($M_n$) $M_n$>25 kDa and polydispersity (PDI) PDI≈2.0.[49] SY has $M_n$>400 kDa and PDI≈3.3. Two different molecular weights of Poly[9,9-bis(6"-(*N*,*N*-dimethyl)-hexyl)-2,7-fluorene]-*alt*-[3,7-dibenzothiophene-*S*,*S*-dioxide] (SPFN) were synthesized by the collaborator. Their molecular weight and polydispersity are around $M_n$=98 kDa with PDI=2.1 and $M_n$= 46 kDa with PDI=2.0.[50,51] Solvents such as anhydrous methanol, toluene, and chlorobenzene were all obtained from Sigma-Aldrich. Molybdenum (VI) oxide 99.95% powder ($MoO_x$) was purchased from Alfa Aesar. Decyltrichlorosilane 97% (DTS) was purchased from Gelest. 200 nm-diamond lapping films were obtained from Allied High Tech Product and Buehler.

**Device Fabrication.** In the fabrication of bilayer Organic Light Emitting Field Effect Transistors (OLEFETs), the bottom transistors were first prepared using diamond lapping films to create grooves on the substrate. The groove dimensions were 50 to 100 nm in width and approximately 10 nm in depth. It is noteworthy that the quality of lapping films has a notable impact on the leakage current in transistors, as demonstrated in the transfer I-V curves shown in Figure S7. Therefore, it is advisable to employ thick dielectric layers >300 nm and to inspect particle sizes in lapping films, as depicted in Figure S8, when high leakage current is observed.

The nano-grooved $SiO_2$ substrates were passivated by DTS. After pssivating DTS, we spin-coated PATBT at 5000 rpm for 40 sec from hot chlorobenzene solution at 90 ℃. Then, a 30 nm-thick PATBT was successfully coated onto the DTS-passivated nano-grooves. Groove density has been shown critical to tune charge carrier mobilities of organic transistors[19]. Therefore, we measured the mobilities for different rubbing times on the substrates as demonstrated in Figure S7a. Based on that, we then chose 4 rubbing cycles and 6 rubbing cycles for two devices with high and low mobilities. The nanogrooved OLEFETs in this work compared to the reference device[52] shows significantly enhanced transport mobilities (0.15 to 0.57 cm$^2$/Vs versus 0.03 cm$^2$/Vs).

For source electrodes on the PATBT, 40 nm gold (Au) were thermally deposited through the silicon shadow mask (See Figure S8). Then we spin-coated the light emitting SY at 2000 rpm 50 sec from a 0.6 wt% solution in toluene to produce a SY layer around 100 nm. After baking the PATBT/SY bilayer at 80 °C for 20 minutes, we spun the conjugated polyelectrolyte SPFN of 0.05 wt% in methanol at 3000 rpm for 50 sec and baked at 70 °C for 10 minutes. For drain electrodes, the drain electrodes were deposited on smooth SPFN in the sequence of Au (1 nm), Ag (7 nm), and MoO$_x$ (30 nm) through the silicon mask. Here Au played a role as a wetting layer to smoothen the following layers. Finally, the 7 nm Ag layer was covered by a MoO$_x$ layer to prevent the hybrid drain from oxidation.

This pair of asymmetric electrodes, the hybrid drain (SPFN/Au:Ag/MoO$_x$) and the source (PATBT/Au), formed a hybrid contact-OLEFET (HC-OLEFET) with a pixel area $0.1 \times 0.0012$ cm$^2$. The transistor channel was fabricated in the 1 mm width and 40 μm length. Commercial highly-doped n-type silicon wafers with dry thermal SiO$_2$ were used as the gate electrodes. The thickness of SiO$_2$ is 300 nm with the corresponding capacitance 10.7 nF/cm$^2$.

**Transmittance of the Hybrid Drains.** To more comprehensively understand the key parameters affecting transmittance of hybrid drains, the Au wetting effect, molecular weight of SPFN, and MoO$_x$ thickness were investigated. We prepared the hybrid drains with/without Au, SPFN/Au:Ag/MoO$_x$ and SPFN/Ag/MoO$_x$, and found that bimetallic contact can guarantee both high electrical conductivity and optical transparency on low-M$_n$ SPFN as described earlier. Furthermore, different MoO$_x$ thicknesses were deposited on the bimetallic contact above both high/low-M$_n$ SPFN. We found that SPFN morphology dominates transmittance when MoO$_x$ thickness <40 nm. Increasing MoO$_x$ thickness on high/low-M$_n$ SPFN shows decreased/increased transmittance, as illustrated in Figure S3. Though thicker MoO$_x$ >40 nm may affect photon extraction, it is unlikely chosen due to the severe degraded conductivity.[53] Therefore, smooth electron injection layers, thin bimetallic contacts, and MoO$_x$ <30 nm are suggested for the optimization of hybrid contact design. Please see more detailed discussion about Figure S3 in the supporting information.

**Materials Characterizations.** The transmission spectra from the multilayer drains with and without 1 nm Au on glass substrates were collected using a reflective spectrometer (Shimadzu UV3600). All the transmission spectra in Figure S2 were normalized to the spectrum of a clear glass substrate. The electroluminescence spectrum of an operating HC-OLEFET at V$_{DS}$=V$_{GS}$=-100 V was collected by a mini-spectrometer (Ocean Optics USB2000+) through a 10× objective, shown in Figure

S2b. The emission collected by the objective goes through a coupling lens on a X-Y translation stage aligning to an optical fiber with 200 µm diameter and NA 0.22, and finally enters a mini-spectrometer for spectra (or a calibrated photomultiplier for brightness[54]). The images of topography and its electronic mapping on the hybrid drains were obtained by the Bruker Innova. Note that the SPFN films must be flushed by methanol to increase smoothness and conductivity before KPFM and cAFM (see the discussion about Figure S5 and S6).

Following the solvent-flushing procedure, we characterized the electronic properties of SY/SPFN samples of KPFM, cAFM, and UV-VIS absorption on the ITO glass substrates 1.5×2.0 cm$^2$. The results revealed a uniform reduction in the barrier to charge injection, a 10-time increase in current, and an observable new spectral feature in the absorption spectra of the SY/SPFN samples. These findings indicate that electronic enhancement was uniformly achieved over a large area. Furthermore, the thickness variation of the flushed SPFN decreased from 11 to 2.2 nm, which is still well within the range of reported thicknesses for ionic salts[15,55] and CPEs[35] (0.5 nm to 15 nm). These results underscore the continued effectiveness of SPFN after solvent-flushing.

To simultaneously measure the electrical and optical performance of HC-LEFETs, we utilized the Keithley source measurement units and a calibrated photomultiplier tube. All HC-OLEFETs were tested in a N$_2$ glovebox with O$_2$ <10 ppm. The reliability of the electrical response of the OLED with identical multilayers in HC-OLEFETs was demonstrated through a continuous ON/OFF test, as presented in Figure S4. Current can be reliably ON/OFF upon voltage switching at 7 V and 10 V though current is decreased by the high-Mn SPFN. Please see more detailed discussion in the SI to avoid the optical and electrical degradation shown in Figure S3 and S4.

**Calculation for the Simulation.** Since the charges transport and interact through dielectric relaxations. Multiplying carrier drift velocity $v = \mu E$ by the dielectric relaxation times $\tau = \epsilon_{SY}/ne\mu$ produces two transit lengths for electrons and holes, $l_n = \epsilon_{SY}E/ne$ and $l_p = \epsilon_{SY}E/pe$. The equation (2) hence can be further converted into a function of p with two fitting parameters n and β as the equation (3) shows,

$$\eta_c = \frac{(1+\frac{\mu_n}{\mu_p})\cdot\frac{\beta n}{p+n}}{1+\frac{n\mu_n}{p\mu_p} - (1+\frac{\mu_n}{\mu_p})\cdot\frac{\beta n}{p+n}} \quad (3)$$

With the contact area 1 mm ×12 µm cm$^2$, the SY thickness 100 nm, and the current value at $V_{DS}=V_{GS}$= -100 V, the maximum hole density $p_{max}$ is roughly=1.1×10$^{16}$ 1/cm$^3$. The maximum electron densities in the two devices $n_{max}$ are about 10$^{18}$ and 10$^{17}$ 1/cm$^3$, respectively found by further substituting $p_{max}$ into the following equation when electrons are 100 % recombined at $\mu_n/\mu_p$=0 demonstrated by the equation (4).

$$\frac{J_c}{J_p} \sim \frac{J_n}{J_p} = \frac{n\mu_n}{p\mu_p} = \frac{1}{7} \text{ and } \frac{1}{16} \quad (4)$$

The $l_n$ is then bounded by 0.18 µm. Since both $l_p$ and $l_n$ are approximately 18 µm and 0.18 µm ≫ 100 nm (the SY thickness), the Coulombic interactions through dielectric responses shall be homogeneous so β is roughly a constant, which matches the extracted β values in Table 2 when $\mu_n/\mu_p$ >4.0×10$^{-4}$ and 1.0×10$^{-3}$ for the device 1 and 2. Before being captured, the injected electrons in a hole-rich environment will undergo diffusion-like transport with diffusion length $\lambda_n = \sqrt{D_n\tau_n} = \sqrt{\epsilon kT/ne^2}$; $\lambda_n$ is about 18.0 nm when n=1.4×10$^{16}$ 1/cm$^3$. Detailed calculation can also be found in the S3 and S4 sections.

## Supporting Information

Details about particle analysis, transmittance, and electroluminescent spectra of the hybrid drains with/without Au; topographic images and transmittance spectra of the hybrid drains composed of different MoO$_x$ thicknesses above high/low-M$_n$ SPFN; Diode I-V curve and ON/OFF reliability test of a OLED with the equivalent layers in HC-OLEFETs; measurement details of KPFM and cAFM on the hybrid drain; KPFM surface potential mapping of SY/SPFN before solvent-flushing; cAFM I-V curves of the SY and SY/SPFN films; the correlation between mobility, EQE, and capture rate; modeling procedures and the applied parameters; the relation between rubbing and transistor performance; Transistor I-V curves with different rubbing cycles and leakage currents; details of asymmetric contact fabrications; SEM images of unqualified and qualified diamond lapping films; photos and zoom-in pictures of 2-inch silicon shadow masks with source and drain patterns; discussion about I$_{DS}$ discrepancy in output and transfer operations; the EQE of an ionic CPE.

## Acknowledgement

The authors are grateful to the financial support of the Ministry of Science and Technology, Taiwan (MOST 110-2112-M-006-025-MY2). We appreciate Dr. Byoung Hoon Lee's and Dr. Jen-Sue Chen's help on the transmission spectrum of the hybrid drain in the supporting information. Dr. Bernard Haochih Liu provided the Bruker Dimension Icon to take a part of cAFM data. Ms. Ting-Yu Huang's effort on the MoO$_x$ topographies in the supporting information is acknowledged. We thank to Dr. Feng-Yin Chang's and Dr. Tzung-Fang Guo's valuable comments.

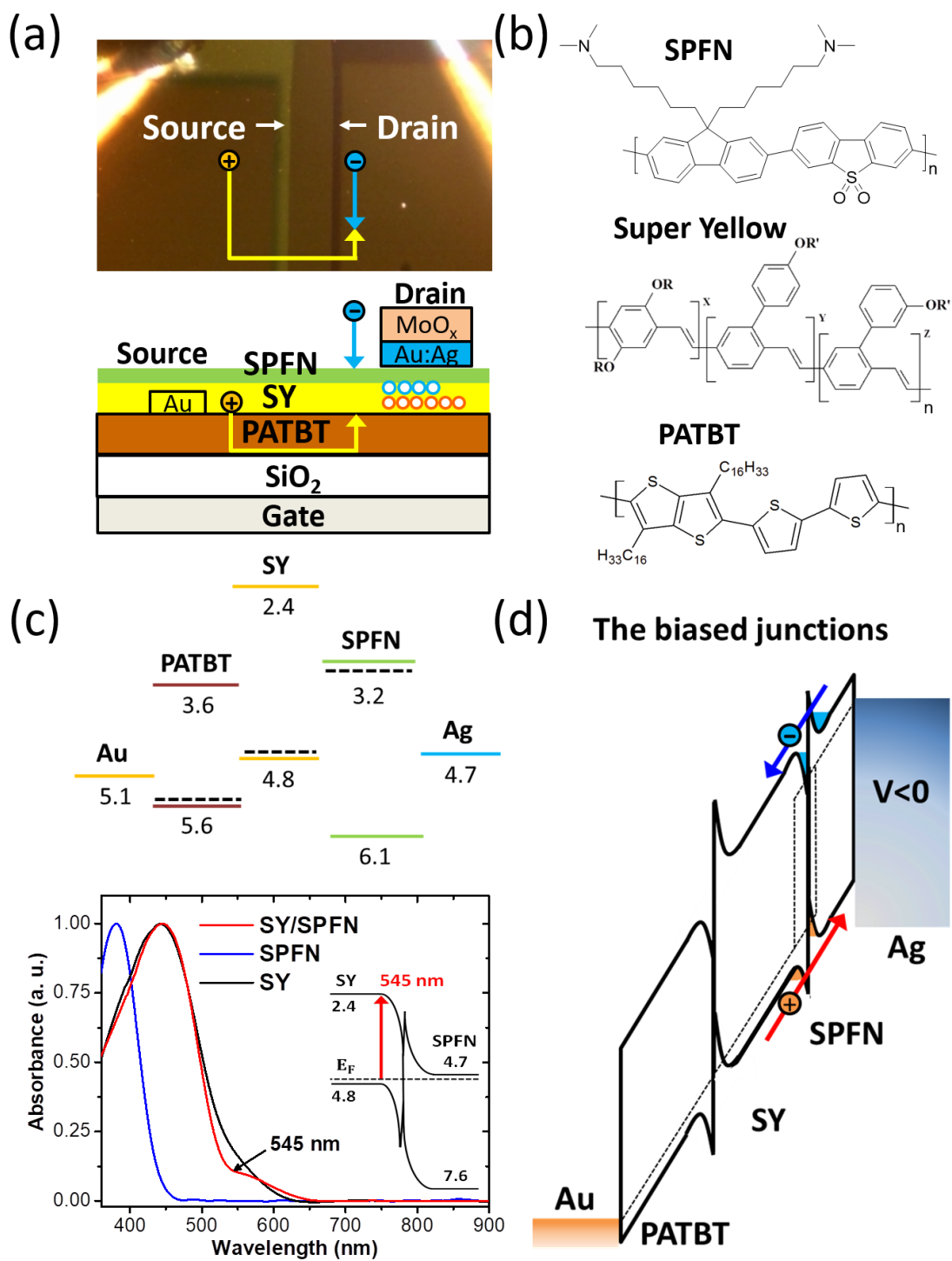

**Figure 1.** (a) HC-OLEFET image and schematic depicting accumulated electrons (blue circles) and holes (orange circles) beneath the hybrid drain, with charge flows indicated by yellow and blue arrows. (b) Chemical structures of n-type conjugated polyelectrolyte (SPFN), Super Yellow (SY), and high-mobility semiconducting polymer (PATBT). (c) Energy diagrams and absorption spectra before (top) and after (bottom) aligning SY and SPFN Fermi levels. (d) Energy diagram during device operation showing accumulated electrons (blue area) and holes (orange area) in the bending potential well

at the light emitting interface.

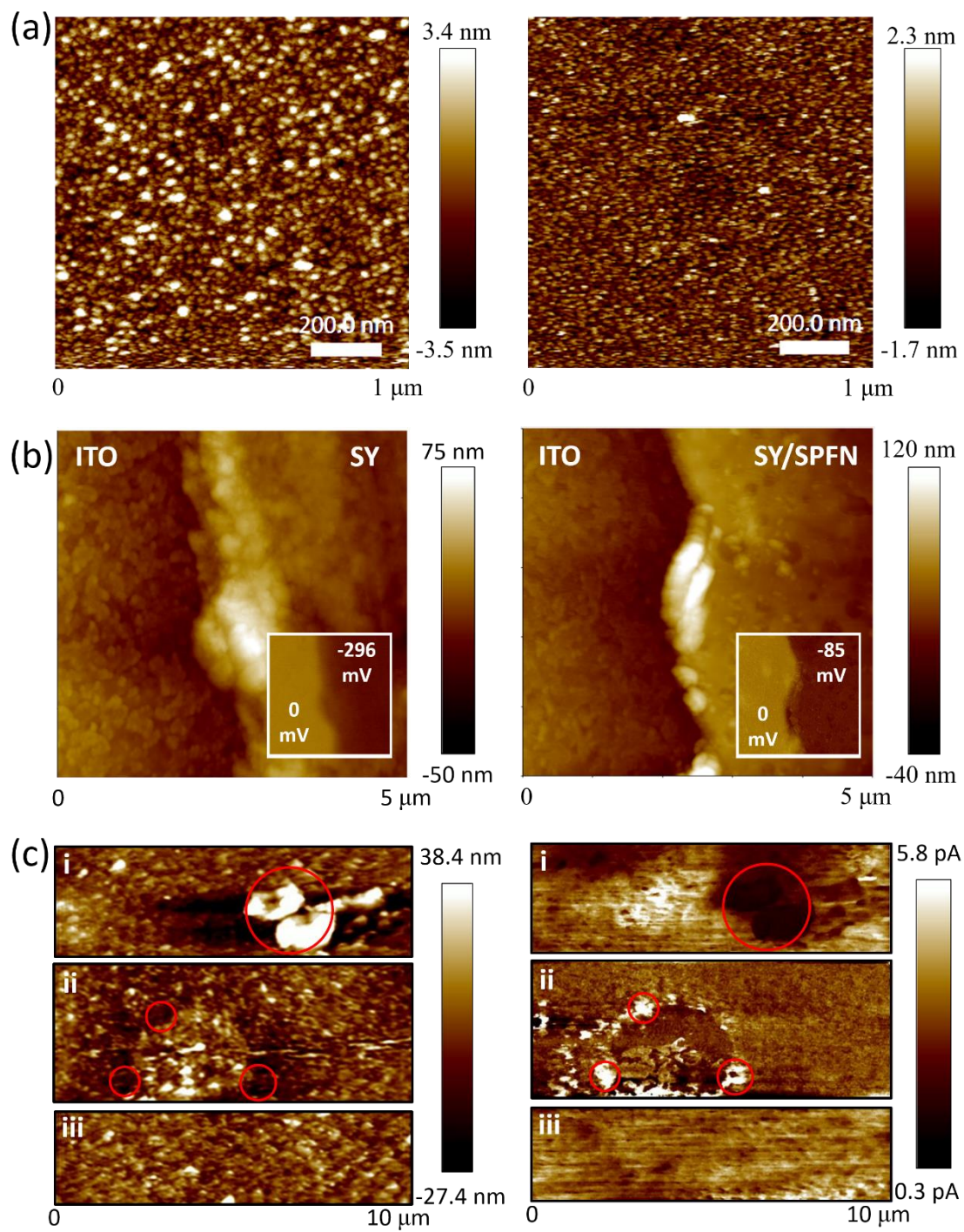

**Figure 2.** AFM images of the hybrid drains. (a) Topography showing the difference in particle size and surface roughness between drains without (left) and with (right) 1 nm Au. Particle size is 10 nm and surface roughness is 1.3 nm for drains without Au, and 4 nm and 0.7 nm for drains with Au. (b) Topography and surface potential mapping for ITO/SY (left) and ITO/SY/SPFN (right). The surface potentials are indicated by insets of $\phi_{SY}$ =-296 mV and $\phi_{SY/SPFN}$ =-85 mV (with reference $\phi_{ITO}$ = 0 mV). (c) Topographic and current mapping of SY/SFPN after 1, 2, and 3 wash cycles. The red

circles indicate the inverse relationship between height and current.

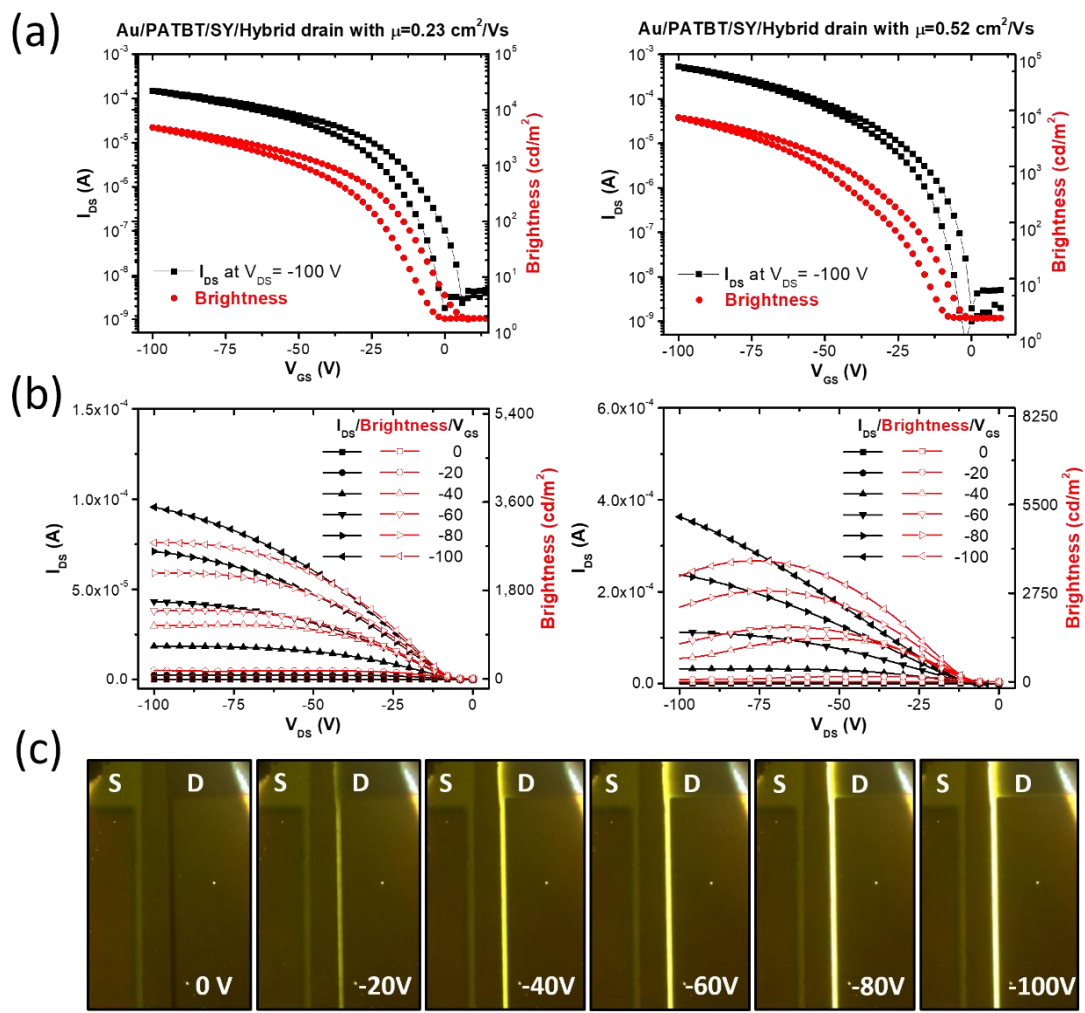

**Figure 3.** Electrical and optical measurements of devices with mobilities of 0.23 cm$^2$/Vs and 0.52 cm$^2$/Vs. (a) Transfer current-voltage (I-V) characteristics at $V_{DS}$= -100 V (b) Output I-V characteristics. (c) Electroluminescence images captured during the output measurements at $V_{DS}$= -100 V and $V_{GS}$= 0 to -100 V in increments of -20 V.

**Table 1.** Electrical and optical performance values collected from the transfer I-V characteristics of Device 1 with 0.23 cm$^2$/Vs and Device 2 with 0.52 cm$^2$/Vs.

| Device | Hole mobility (cm$^2$/Vs) | On/off ratio | Current density (mA/cm$^2$) | Brightness (cd/m$^2$) | EQE (%) |
|---|---|---|---|---|---|
| 1 | 0.23 | $3 \times 10^4$ | 0.9 – 750 | 7 – 4800 | 0.51 |
| 2 | 0.52 | $1 \times 10^5$ | 1.2 – 2700 | 4 – 7700 | 0.23 |

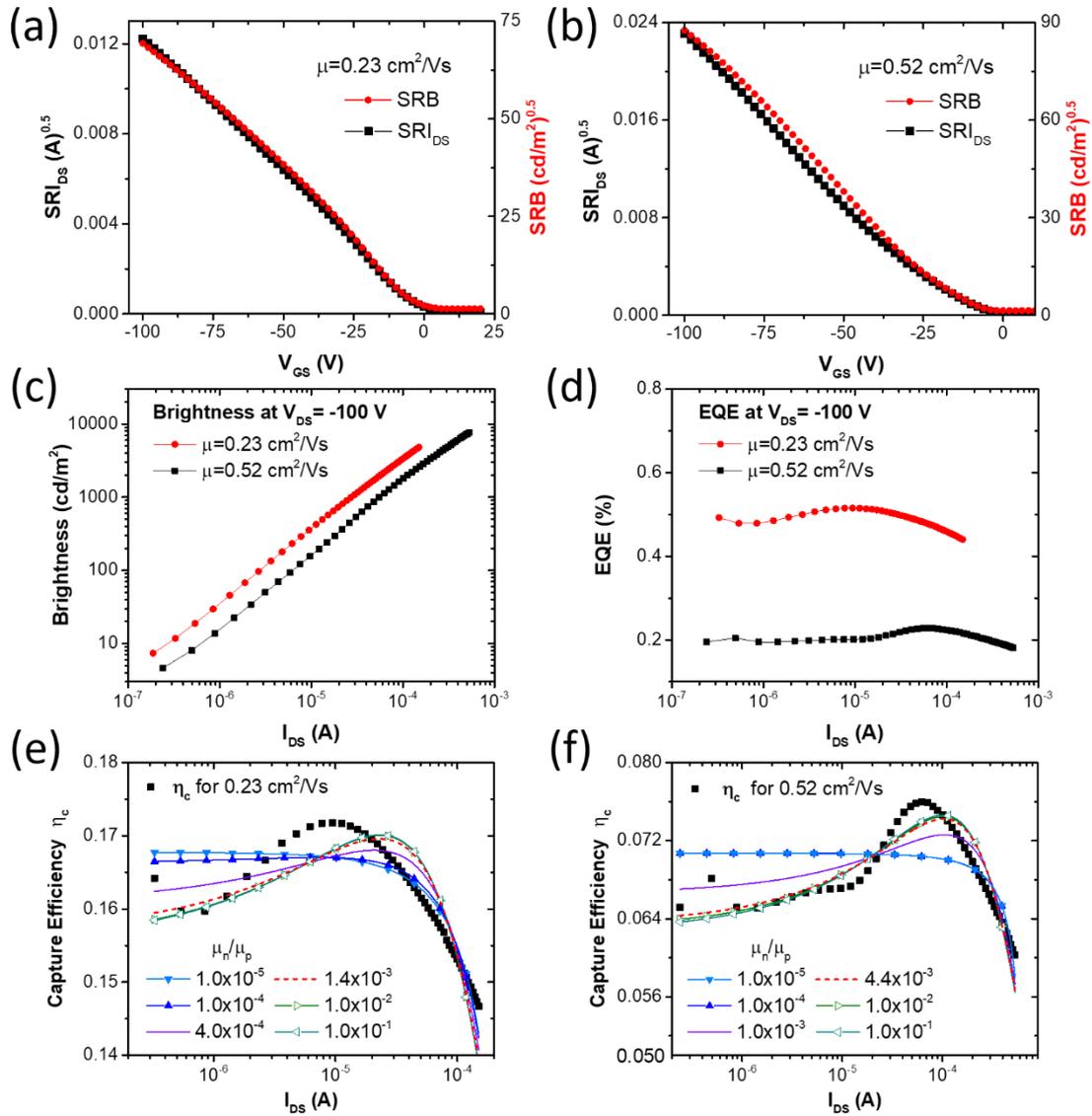

**Figure 4.** Characterization of devices with μ=0.23 cm$^2$/Vs and 0.52 cm$^2$/Vs. (a) and (b) Plots of the square root of $I_{DS}$ (SRI$_{DS}$, black solid square) and brightness (SRB, red solid circle) versus $V_{GS}$. (c) Brightness vs. $I_{DS}$ curves for both devices. (d) EQE vs. $I_{DS}$ at $V_{DS}$= -100 V. (e) and (f) Simulated capture efficiency $\eta_c$ at various $\mu_n/\mu_p$ values (colored curves) with the experimental data (black solid squares). Purple curves distinguish two capture dynamics represented by solid and open triangles. The red-dashed curves represent the simulated $\eta_c$ with n=1.4×10$^{16}$ 1/cm$^3$ and corresponding diffusion length $\lambda_n$=18 nm.

**Table 2.** Summary of extracted n and β values from simulated $\eta_c(I_{DS})$ curves for Device 1 and 2 with various mobility ratios $\mu_n/\mu_p$.

| $\mu_n/\mu_p$ | | $10^{-5}$ | $10^{-4}$ | $4\times10^{-4}$ | $1.4\times10^{-3}$ | $10^{-2}$ | $10^{-1}$ |
|---|---|---|---|---|---|---|---|
| Device 1[a] | n | $2.4 \times 10^{19}$ | $2.8 \times 10^{18}$ | $9.7 \times 10^{16}$ | $1.4 \times 10^{16}$ | $1.7 \times 10^{15}$ | $1.7 \times 10^{14}$ |
| | β | $1.7 \times 10^{-6}$ | $1.7 \times 10^{-5}$ | $4.0 \times 10^{-5}$ | $4.2 \times 10^{-5}$ | $4.3 \times 10^{-5}$ | $4.2 \times 10^{-5}$ |
| $\mu_n/\mu_p$ | | $10^{-5}$ | $10^{-4}$ | $10^{-3}$ | $4.4\times10^{-3}$ | $10^{-2}$ | $10^{-1}$ |
| Device 2[a] | n | $1.9 \times 10^{20}$ | $1.9 \times 10^{19}$ | $1.2 \times 10^{17}$ | $1.4 \times 10^{16}$ | $5.6 \times 10^{15}$ | $5.3 \times 10^{14}$ |
| | β | $7.1 \times 10^{-7}$ | $7.1 \times 10^{-6}$ | $3.9 \times 10^{-5}$ | $3.9 \times 10^{-5}$ | $3.9 \times 10^{-5}$ | $3.6 \times 10^{-5}$ |

[a] Device 1 has a mobility 0.23 cm$^2$/Vs and Device 2 has a mobility 0.52 cm$^2$/Vs.

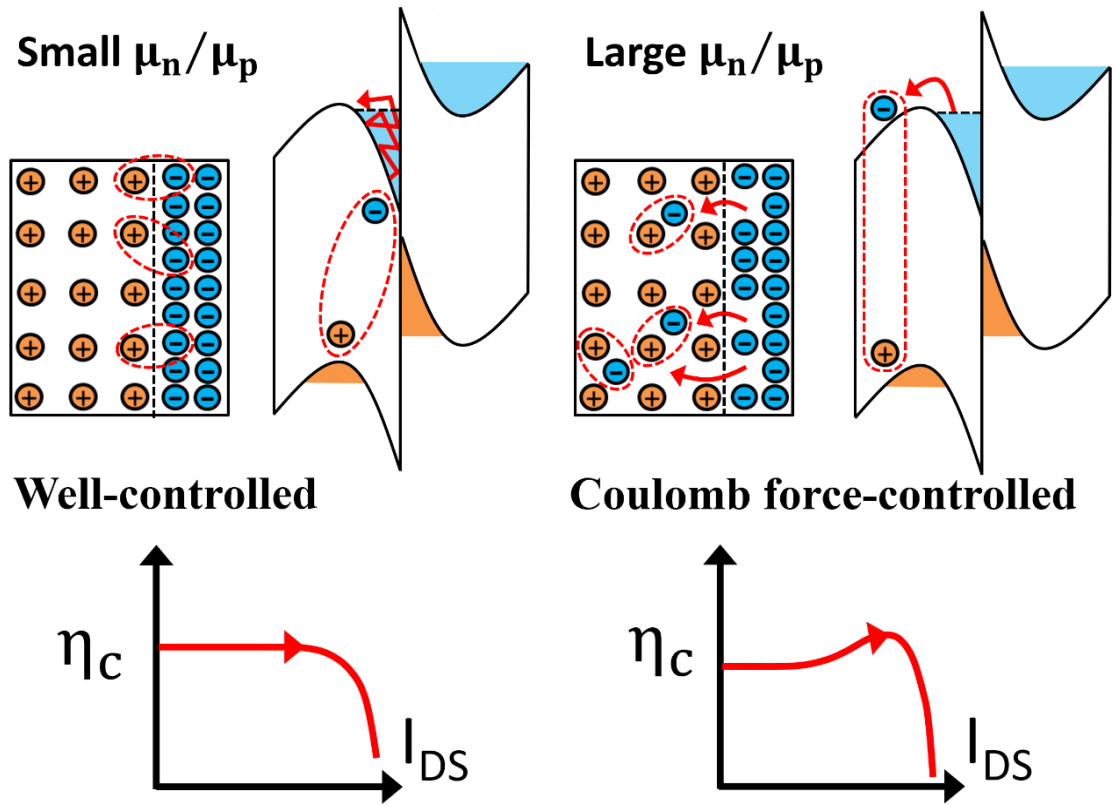

**Figure 5** Two different capture dynamics based on accumulated and escaped electrons. When the ratio of electron mobility to hole mobility $\mu_n/\mu_p$ is low, electrons are trapped and the capture efficiency $\eta_c$ follows a kinetic process controlled by the potential well. On the other hand, when $\mu_n/\mu_p$ is high, the transport and capture of escaped electrons are restricted by the thermal capture radius of Coulomb forces.

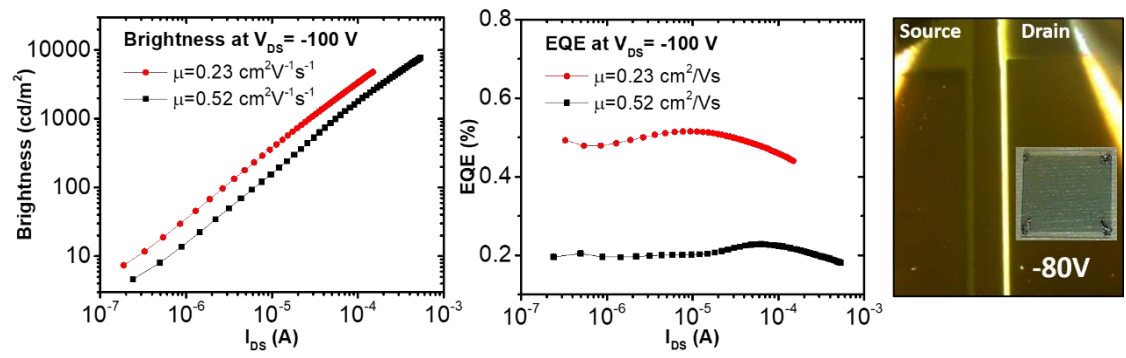

Table of Contents graphic